\documentclass[floats,floatfix,showpacs,amssymb,prd,twocolumn,superscriptaddress,nofootinbib]{revtex4-1}

\usepackage{amssymb,amsmath,verbatim,mathtools,needspace,etoolbox,graphicx,physics,microtype,afterpage,bm}

\usepackage[dvipsnames, usenames]{xcolor}
\usepackage[english]{babel}
\usepackage{lmodern}
\usepackage{booktabs}
\usepackage{amsfonts}
\usepackage{listings}
\usepackage{fancyhdr}
\usepackage{xcolor}
\usepackage{comment}
\usepackage{enumerate}
\newcommand{\olu}{\mathcal{O}_{^U}^{^L}}
\newcommand{\blu}{\mathcal{B}_{^U}^{^L}}
\newcommand{\blus}{{\mathcal{B}_{^U}^{^L}}^*}
\newcommand{\plu}{\mathcal{P}_{^U}^{^L}}

\newcommand{\hu}{\mathcal{H}_U}
\newcommand{\hl}{\mathcal{H}_L}
\newcommand{\ho}{\mathcal{H}_0}
\newcommand{\bd}{\bm{d}}
\newcommand{\bt}{\bm{\theta}}
\newcommand{\btu}{\bm{\theta}_\mathrm{U}}
\newcommand{\bteff}{\bm{\theta}_{\mathrm{eff}}}

\definecolor{linkcolor}{rgb}{0.0,0.3,0.5}
\usepackage[unicode, colorlinks=true, linkcolor=linkcolor, citecolor=linkcolor, filecolor=linkcolor,urlcolor=linkcolor, pdfusetitle]{hyperref}
\usepackage[all]{hypcap}
\usepackage[T1]{fontenc}
\usepackage[utf8]{inputenc}
\usepackage{tabularx}
\usepackage{float}
\allowdisplaybreaks
\graphicspath{{./figures/}}

\def\apj{\rm{ApJ}}

\def\mnras{\rm{MNRAS}}
\def\prd{\rm{PRD}}
\def\prl{\rm{PRL}}

\newcommand{\cuhk}{\affiliation{Department of Physics, The Chinese University of Hong Kong, Shatin, N.T., Hong Kong}}
\newcommand{\umich}{\affiliation{Physics Department, University of Michigan, Ann Arbor, MI 48109, USA}}
\newcommand{\pisa}{\affiliation{Dipartimento di Fisica ``E. Fermi", Università di Pisa, I-56127 Pisa, Italy}}
\newcommand{\lit}{\affiliation{Laboratoire des 2 Infinis - Toulouse (L2IT-IN2P3), Université de Toulouse, CNRS, UPS, F-31062 Toulouse Cedex 9, France}}
\newcommand{\unimib}{\affiliation{Dipartimento di Fisica “G. Occhialini”, Università degli Studi di Milano-Bicocca, Piazza della Scienza 3, 20126 Milano, Italy}}
\newcommand{\infn}{\affiliation{INFN, Sezione di Milano-Bicocca, Piazza della Scienza 3, 20126 Milano, Italy}}

\definecolor{rb4}{HTML}{27408B}

\begin{document}

\title{Mitigating the effect of the population model uncertainty on the strong lensing Bayes factor using nonparametric methods}

\pacs{}

\author{Damon H. T. Cheung} 
\email{damoncht@umich.edu}
\cuhk
\umich

\author{Stefano Rinaldi} 
\email{stefano.rinaldi@phd.unipi.it}
\pisa

\author{Martina Toscani} 
\lit
\unimib
\infn
\author{Otto A. Hannuksela} 
\cuhk

\date{\today}
\begin{abstract}
Strong lensing of gravitational waves can produce several detectable images as repeated events in the upcoming observing runs, which can be detected with the posterior overlap analysis (Bayes factor). 
The choice of the binary black hole population plays an important role in the analysis as two gravitational-wave events could be similar either because of lensing or astrophysical coincidence.
In this study, we investigate the biases induced by different population models on the Bayes factor. 
We build up a mock catalog of gravitational-wave events following a benchmark population and reconstruct it using both nonparametric and parametric methods. 
Using these reconstructions, we compute the Bayes factor for lensed pair events by utilizing both models and compare the results with a benchmark model.
We show that the use of a nonparametric population model gives a smaller bias than parametric population models.
Therefore, our study demonstrates the importance of choosing a sufficiently agnostic population model for strong lensing analyzes.
\end{abstract}
\maketitle

\section{Introduction}

General relativity posits that massive objects bend spacetime in a way that focuses the light rays that travel near them and causes gravitational lensing \citep[][]{Schneider1992GravitationalLenses,Bartelmann_2010}. 
Through decades of technological innovation, gravitational lensing has grown from an abstract theoretical idea into a standard tool in the study of the cosmos. 
For example, it has historically provided among the most stringent evidence for the existence of dark matter~\citep{Clowe_2004,Markevitch_2004}, aided in exoplanet discoveries~\citep{Bond2004OGLEEvent}, and allowed scientists to measure the expansion of the Universe~\citep{time_delay_cosmo}. 

Similar to electromagnetic waves, gravitational waves (GWs) can also be gravitationally lensed if they travel near massive astrophysical objects \citep{focusiing_Gradiation,Wang:1996as,nakamura1999wave,Wave_effects_lensing}. 
Such lensing of GWs could produce a plethora of observational signatures. 
In the so-called geometrical optics limit,\footnote{The geometrical optic limit is the situation in which the GW wavelength is significantly smaller than the Schwarzschild radius of the lens.} lensing of GWs is similar to traditional lensing of light~\citep{Takahashi2003WaveBinaries}. 
In this limit, lensing can  
\begin{enumerate}[(i)]
\item magnify the GW~\citep{Pang2020LensedDetection}, producing an apparently high-mass distribution of binary black holes (BBHs)~\citep{Dai2016EffectMergers,Oguri2018EffectMergers,Broadhurst2018ReinterpretingDistances,Cusin2021LensingDetectors} that might be detectable if the source is a neutron star merger~\citep{Pang2020LensedDetection},
\item invert the GW along one or both principal axes~\citep{Dai2017OnWaves,Ezquiaga2020PhaseWaves} that may be detectable as a discernible complex phase shift of waveform with higher order modes~\citep{Dai2017OnWaves,Ezquiaga2020PhaseWaves,Wang2021IdentifyingDetectors,Janquart2021OnModes,vijaykumar2022detection},
\item split the signal into multiple images detectable as repeated events in the detectors~\citep{Haris2018IdentifyingMergers,Hannuksela2019SearchEvents,Li2019FindingEvents,Broadhurst2019TwinMerger,McIsaac2019SearchRuns,Liu2020IdentifyingVirgo,Dai2020SearchO2,TheLIGOScientificCollaboration2023SearchRun,Kim2022Deep-2}.
\end{enumerate}
In the wave optics limit,\footnote{The wave optics limit is the situation in which the GW wavelength is comparable to or larger than the Schwarzschild radius of the lens.} the GW can experience frequency-dependent deformations that might be captured with template-based or machine-learning searches~\citep{cao2014gravitational,Lai2018DiscoveringLensing,Christian2018DetectingObservatories,Dai2018DetectingWaves,Singh2018ClassifyingLearning,Kim2020IdentificationLearning,Cheung2020Stellar-massWaves,Yeung2021MicrolensingMacroimages,Bulashenko2021LensingPattern,Seo2021StrongMicrolensing,Tambalo2022LensingLenses}. 
Thus far, there have been a number of searches for these GW lensing effects in LIGO-Virgo-Kagra (LVK) data~\citep{Hannuksela2019SearchEvents,Li2019FindingEvents,McIsaac2019SearchRuns,Liu2020IdentifyingVirgo,Dai2020SearchO2,TheLIGOScientificCollaboration2023SearchRun,Kim2022Deep-2,o12opendata,theligoscientificcollaboration2023open}, but there has been no widely accepted detection, despite a recent reinterpretation of the LVK lensing search results~\citep{Diego2021EvidenceLIGO-Virgo}. 

If lensed GW events are observed, they could enable many scientific pursuits, including the detection of intermediate-mass and primordial BHs \citep{Lai2018DiscoveringLensing,Diego_2020,Basak2022ConstraintsMicrolensing},  MACHOs~\citep{Basak2022ConstraintsMicrolensing}, and a better comprehension of the nature of dark matter~\citep{Basak2022ConstraintsMicrolensing,Oguri2020ProbingWaves,Oguri2022AmplitudeLensing,Wang2021LensingObservations,Cao2022DirectWaves}. 
The effectively expanded detector network that can detect lensed events and has excellent timing accuracy may also enable the search for the associations of fast radio bursts and GWs \citep{FRBdejavu}, precise tests of the cosmology \citep{jana2022cosmography}, speed of gravity
\citep{Collett2016TestingLensing,Fan2017SpeedSignals,Baker2016Multi-MessengerWaves,Ezquiaga2022ModifiedLensing}, polarization~\citep{Goyal2020TestingSignals}, and propagation~\citep{Mukherjee2019Multi-messengerWaves,Mukherjee2019ProbingSurveys,Chung2021LensingMass,Finke2021ProbingBinaries,Iacovelli2022ModifiedFunction}. 
In the wave optics limit, they may break the mass-sheet degeneracy~\citep{Cremonese2021BreakingEvents} and allow to measure the velocity of the lens~\citep{Itoh2009AWaves}. 
When strongly lensed, they might allow for subarcsecond precision localization of merging BHs~\citep{Hannuksela2020LocalizingLensing,Wempe2022AObservations}, precision tests of cosmology~\citep{Sereno2010StrongLISA,Liao2017PrecisionSignals,Hannuksela2020LocalizingLensing,Cao2021ALensing}, and combined studies of lensed binaries in the GW and electromagnetic band~\citep{Smith2022DiscoveringObservatory}. 

Strong lensing has been of particular interest in recent years. 
It has been forecasted to become detectable at design detector sensitivity, assuming that the current BBH detections originate from stellar merger remnants~\citep{Ng2018PreciseHoles,,Oguri2018EffectMergers,Xu2021PleasePopulations,Wierda2021BeyondLensing,Smith2022DiscoveringObservatory}\footnote{Observational limits on the rate of lensing have also been set, agnostically, based on the nondetection of the stochastic GW background~\citep{Buscicchio2020ConstrainingBackground,Buscicchio2020ConstrainingBackgroundb,Mukherjee2020InferringBackground,TheLIGOScientificCollaboration2023SearchRun}.}. 
Consequently, there has been a major effort to develop the necessary tools to detect and analyze strong lensing~\citep{Haris2018IdentifyingMergers,Dai2020SearchO2,Liu2020IdentifyingVirgo,Cheung2020Stellar-massWaves,Hannuksela2020LocalizingLensing,Seo2021StrongMicrolensing,Lo2021ASignals,Janquart2021AEvents,Janquart2021OnModes,Goyal2021RapidLearning,Janquart2022OrderingCapabilities,goyal2023rapid}. 

\begin{figure*}
    \centering
    \includegraphics[scale=0.7]{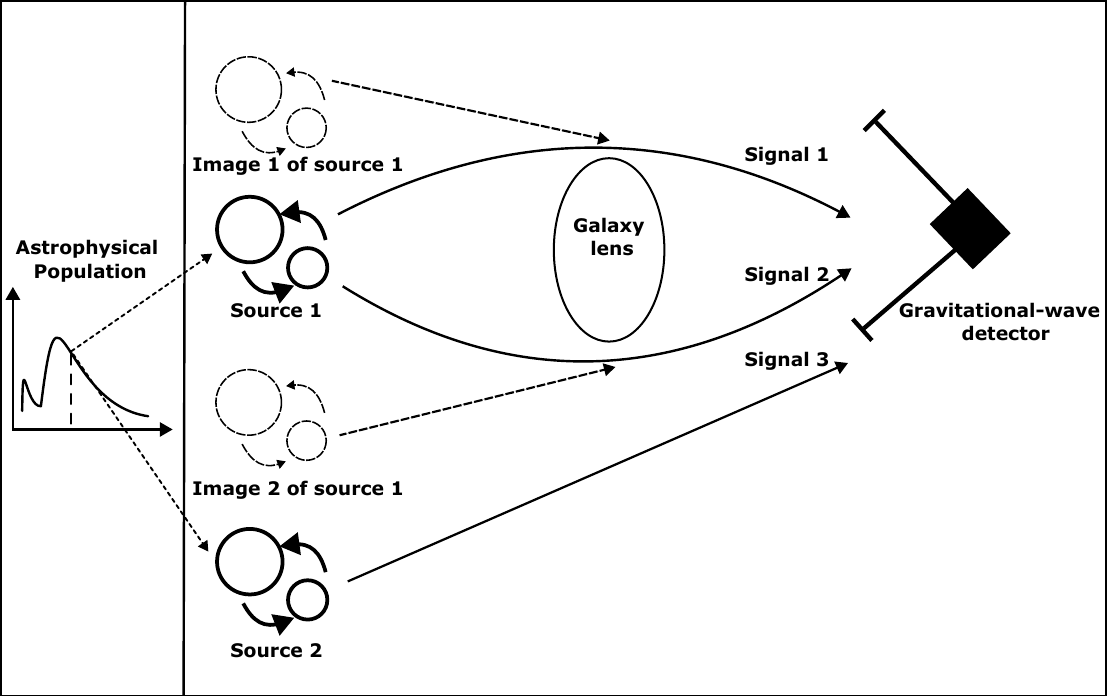}
    \caption{ 
    A strong lens can produce multiple images of a binary black hole (BBH). 
    Although we cannot discern the images in the sky with a ground-based gravitational-wave detector due to poor angular resolution, we can detect them as “repeated events”---two or more gravitational-wave copies that only differ in their amplitudes, phases, and overall arrival times, but are otherwise identical. 
    Indeed, to detect strong lensing, we look for multiple copies of the same event. 
    However, due to finite detector accuracy, a nonlensed BBH may also appear to be similar to other events by astrophysical coincidence that mimics lensing, as indicated in the diagram. 
    The probability of finding such lensing mimickers depends on the underlying population of BBHs. 
    Correctly modeling the BBH population on lensed detections is therefore vital to detections.}
    \label{fig:Illustration}
\end{figure*}

The current strong lensing analysis methods all search for repeated events: 
two or more GW events that appear to have the same frequency evolution and originate from the same sky location, in line with the strong lensing expectation. 
These waves would have identical detector-frame parameters and sky location but differ in arrival times and luminosity due to magnification after strong lensing and overall phases as they pass through caustics~\citep{Takahashi2003WaveBinaries,Dai2017OnWaves}.
Thus, in practice, a strong lensing analysis inspects how similar/dissimilar the detector-frame parameters and sky location are between two or more GW events. 
Alternatively, one could state that a strong lensing search attempts to identify two or more copies of binary compact object systems that appear to be near-identical. 
However, in more technical terms, the analyzes evaluate the strong lensing Bayes factor---the evidence ratio between lensed and unlensed hypothesis. 
The Bayes factor quantifies the support for an event that is strongly lensed versus only appearing to be strongly lensed due to astrophysical coincidence (see Fig.~\ref{fig:Illustration}), and is therefore of utmost importance for strong lensing detections.

Indeed, the probability of encountering strong lensing mimickers increases quadratically as the number of detections increases, a challenge that could be mitigated by higher signal-to-noise ratios or the inclusion of lensing time-delay estimates which however necessitate the use of accurate lens population models. 

Another challenge associated with strong lensing detections is their reliance on population modeling. 
In particular, the likelihood that two or more near-identical binary compact objects appear in the GW detector through astrophysical coincidence depends on the population model: 
if the Universe produces many similar compact binaries, then the chance of them being an astrophysical coincidence is high, and vice-versa. 
There has been a study on the degree to which unlensed events mimic
lensed ones because of the overlap of parameters due to random coincidence~\citep{lensing_or_luck}. However, the statistic used does not include population terms that account for the probability of coincidence.  
To make strong lensing detections, one must consider the probability of astrophysical coincidence in the lensing statistic. 
Therefore, one must have an accurate BBH population model. 
This representation can, in principle, be inferred from LVK data directly: however, one must be careful to use a model that is able to capture all the features that might be present in the population.
Currently, phenomenological parametric population fits, motivated by physical arguments but not directly linked to physical processes, are used, which might in principle capture the characteristics of the source population, \textit{if} the model is sophisticated enough.
However, there has not yet been a significant investigation to understand to what extent the use of these parametric models could bias strong lensing detections. 

Thus, here we investigate the following two questions: 
\begin{enumerate}[(a)]
\item Could the utilization of an incorrect parametric population model lead to an inaccurate strong lensing analysis? 
\item If so, can we overcome this challenge using nonparametric population models?  
\end{enumerate}
In particular, the use of different GW population models in the analysis will affect the prediction and can result in a wrong classification (false detection of lensing). 

In this study, we build up a mock catalog of GW events (benchmark population model) that mimic the preferred model from \citep{o3pop}, the \textsc{PowerLaw+Peak} model (PP model, thereinafter), and generate the posterior samples of the observed events. On top of this catalog, we simulate also a set of lensed pairs of events.
Then, we calculate the Bayes factor of the lensed pairs based on four population models:
\begin{enumerate}[(i)]
\item Benchmark population model that mimics PP model;
\item A hierarchy of Dirichlet process Gaussian mixture models ((H)DPGMM) - Nonparametric population model inferred from the mock catalog;
\item Power-law model---Parametric population model inferred from the mock catalog;
\item Uniform model---events are uniformly distributed across the parameter space.
\end{enumerate}
We will then show  how the Bayes factor of lensed pairs changes with these different population models.

This paper is structured as follows: in Sec.~\ref{sec:lensing}, we review the strong lensing mechanism of GW events and posterior overlap analysis using Bayes factor. 
In Sec.~\ref{sec:population}, we give the details of the population models we use in our analysis, both in the simulation and in the recovery. 
Sec.~\ref{sec:results} illustrates the performance of the various population models in lensing detections. 
In Sec.~\ref{sec:conclusions}, we discuss the results and conclude, arguing that the use of nonparametric population models is suitable for strong lensing detections, while the use of parametric population models may lead to biases.

\section{Gravitational-wave lensing}\label{sec:lensing}
If a BBH is strongly lensed by a galaxy or a galaxy cluster, GWs would split into multiple ``images'', observable as repeated near-identical\footnote{Differing only in amplitude, arrival time, and overall phase.} copies of the same binary. 
If, in addition, the GW encounters smaller lenses (such as stars, primordial BHs, globular clusters, or dark matter subhalos) along its travel path, they can induce frequency-dependent beating patterns and wave optics modulations (often referred to as GW microlensing and millilensing). 
In the limit of weak lensing, the GW would merely be magnified, and the resulting weakly lensed GW would have a different amplitude.
In this work, we focus on strong lensing of binary stellar remnants by galaxies and galaxy clusters, and neglect the effect of micro-, milli- and wave optics lensing. 

In the strong lensing scenario, that is, when the binary is sufficiently aligned with a galaxy or a galaxy cluster such that strong lensing occurs, the lens produces several images of the BBH. 
More specifically, the strong lens produces $2N+1$ images of the source (usually 2 or 4 visible images for a galaxy lens); each image would be related to the others by some magnification $\mu$, lensing time delay $t_d$, and Morse phase $n$, such that the $i$th GW strain~\citep{Takahashi2003WaveBinaries}
\begin{equation}\label{eq:lensed_strain}
    h_L^{i}(f) = |\mu_i|^{1/2} e^{i (2 \pi f t_i^d - \pi n_i \text{sign}(f))} h(f;\theta)\,,
\end{equation}
where $h(f;\theta)$ is the GW strain in the absence of strong GW lensing, $\theta := \{m_1,m_2,\vec{s}_1,\vec{s}_2, z_s, \theta_{J_N}, \alpha, \delta, \psi, t^{\rm u}\}$,\footnote{$m_{1,2}$ denote the component masses, $\vec{s}_{1,2}$ the BH spins, $z_s$ the source redshift, $\theta_{J_N}$ the inclination between total angular momentum $J$ and the direction of propagation, $\alpha$ and $\delta$ the right ascension and declination, $\psi$ the polarization angle, and $t^{\rm u}$ the arrival time of the emitted GW at the detector} are the BBH source parameters. 
The Morse phase $n_i$ equals $0$, $1/2$, and $1$ for minima, saddle, and maxima points of the lensing time-delays and can be solved when the lens geometry is known. 

Note that although it is common to use a strong lensing model to evaluate the image properties $\nu_i$ (magnifications, time delays, and Morse phases), it is typically more convenient for us to work with the image properties directly. 
Even more convenient is working with so-called effective parameters. 
In particular, absorbing the magnification into an “effective” luminosity distance of the $i$th image, 
\begin{equation} \label{eq:magnified luminosity distance}
    d_L^{\rm eff,i}=d_L/|\mu_i|^{1/2}\,,
\end{equation} and the lensing time-delay into the effective arrival time of the $i$th image 
\begin{equation}
    t_i=t_0+t_i^d\,,
\end{equation} is more convenient from an analysis perspective.
By using effective parameters, one can model the two lensing effects using a general-relativistic waveform without modifying anything but the luminosity distance and arrival time terms. 
I.e., we can absorb the lensing modification and time delay by simply replacing all luminosity distance and arrival time variables with their effective counterparts. 
Such a procedure is attractive because it does not require changing existing analysis infrastructure.\footnote{Similar to many prior works, here we ignore the Morse phase term, which could be absorbed in the coalescence phase when the GW consists primarily of (2,2) modes; the inclusion of the Morse phase would merely add a phase term into the analysis that might allow for a slightly improved detection criterion~\citep{Dai2017OnWaves,Ezquiaga2020PhaseWaves,Vijaykumar2022DetectionSignals}.}

\subsection{Lens modeling}

With the form of the strongly lensed GW strain (Eq.~\ref{eq:lensed_strain}) known, one can model strongly lensed GWs, given that we know the image properties.
However, it is still important to ask the question of how one obtains the image properties in the first place. 
These can be solved from the lens equation\footnote{We refer those interested in relating the image parameters $\nu$ to the lens model to standard lensing textbooks such as~\citep{Congdon2018PrinciplesLensing}.}, given a strong lens model. 
It was realized early on that relatively simple \emph{smooth} lens mass models\footnote{E.g., power-law ellipsoidal profiles with shear.} fit the strong galaxy lens flux observations on large scales quite well~\citep{Mao1997EvidenceGalaxies,Metcalf2001CompoundHalos}, while modeling substructure within the lens and along the line-of-sight can be captured relatively well with a Gaussian uncertainty term to the magnification~\citep[e.g.][and references therein]{Wempe2022AObservations}. 
Such galaxy lens models could be used for data-analysis purposes. 
However, for the purposes of mock data and benchmarking, simpler lens models are often favored, as they can be used to produce data relatively quickly without complex lens morphology and population modeling. 
Since we are primarily interested in the source population modeling and not the lens modeling, we opt for this latter approach.

In particular, we assume that the \emph{a priori} distribution of magnifications for each GW image, $\mu$, follows a power-law distribution~\citep{Narayan1996LecturesLensing}
\begin{equation}\label{eq:magnification_marginalised}
    p(\mu) \propto \mu^{-3}\,,
\end{equation}
which is appears to be correct near the caustic (i.e., when the magnification is high, $\mu \gtrsim 10$).
Based on simulations and forecasts for the current and near future detector network, most detected GW events are relatively highly magnified~\citep{Dai2016EffectMergers,Ng2018PreciseHoles, Xu2021PleasePopulations,Wierda2021BeyondLensing,Smith2022DiscoveringObservatory}, which justifies the high-magnification limit. 

While the magnification distribution of any given image follows the distribution above, the magnifications between the images are correlated. 
Specifically, in most strong lensing configurations where the signal is heavily magnified, the source is close to a caustic, and due to near-symmetry typical strong lenses, the magnifications of both images are comparable ($\mu_1\sim \mu_2$, \citealt{More2021ImprovedEvents})\interfootnotelinepenalty=10000\footnote{For the singular isothermal sphere (SIS) model, $\mu_1 = \mu_2$ in the limit $\mu_1\rightarrow \infty$. A similar result can be derived for the singular isothermal ellipsoid (SIE) models.}.
However, the magnifications are rarely precisely the same, except in highly simplified lens models. 
Assuming that the magnification ratios were exactly known would produce highly optimistic results. 
Thus, to correct for lens modeling uncertainties, including ellipticities, shears, and substructure, one might assume, as a simple approximation, that the distribution of the second magnification follows 
\begin{align}\label{eq:magnification_correlated}
    p(\mu_2|\mu_1)\propto \mathcal{N}(\mu_{2,\mathrm{SIS}},0.15 \cdot  \mu_{2,\mathrm{SIS}})\,,
\end{align}
where \(\mu_1\) is the magnification for the first image, \(\mu_{2,\mathrm{SIS}}\) is the magnification for the second image calculated using a SIS model, \(\mu_2\) is the final magnification for the second image drawn from a normal distribution \(\mathcal{N}(\mu, \sigma)\) with mean \(\mu\) and standard deviation \(\sigma\).
Such a spread in the magnification induced by microlensing and the substructure is conservative for high-signal events when neglecting the nonlinear effect and is used in many prior studies~\citep{Hannuksela2020LocalizingLensing,Wempe2022AObservations}, and it can generally be regarded as better than presuming that the magnifications were fixed, for example, by a singular isothermal sphere lens model.

\subsection{Strong lensing Bayes factor}

The quantity we will use to decide whether a pair of events is likely to be lensed or not is the strong lensing Bayes factor $\blu$.
The strong lensing hypothesis states that the set of events originates from the same astrophysical event. 
Therefore, each of the detector-frame parameters will be the same under this hypothesis~\citep{Haris2018IdentifyingMergers}. 
The unlensed hypothesis states that the set of signals was produced by unrelated astrophysical events (in this case, compact binary coalescences). 

In principle, the strong lensing Bayes factor should include all detector-frame binary parameters and lens parameters~\citep{Liu2020IdentifyingVirgo,Lo2021ASignals,Janquart2021AEvents}. 
Evaluating the Bayes factor would require constructing and estimating the joint likelihood between the set of waveforms. 
However, such an evaluation is highly demanding from a computational point of view and thus impossible for us to use at high throughput when performing an analysis at the population level.

Therefore, we opt to use a well-developed and widely used technique, called \emph{posterior overlap analysis}, to estimate a ``subset'' of the Bayes factor, used for example in~\citep{Haris2018IdentifyingMergers}.
In this study, we focus on lensing that produces 2 images with the subset of the binary parameters $\bt^\prime$ are $m_1^z,m_2^z$,
which correspond to the detector-frame primary mass and secondary mass are unaffected by the gravitational lensing.
We marginalize over $\bt^\prime$ to get the following strong lensing Bayes factor (see Appendix \ref{A:blu_derivation} for the full derivation): 
\begin{equation} \label{eq:blu evidence ratio}
\begin{aligned} 
\blu    &= \frac{p(\bd|\hl)}{p(\bd|\hu)}= \frac{p(d_1,d_2|\hl)}{p(d_1|\hu)p(d_2|\hu)},
\end{aligned}
\end{equation}
where $p(d_1,d_2|\hl)$ is the evidence of having data $d_1,d_2$ under lensed hypothesis and $p(d_i|\hu)$ is the evidence of having data $d_i$ under unlensed hypothesis. The Bayes factor (Eq.~\ref{eq:blu evidence ratio}) can then be expanded in binary parameter space as  
\begin{equation} \label{eq:blu formula}
\begin{aligned} 
\blu        &=\dfrac{\int  \dfrac{p_{\mathrm{obs}}(\bt^\prime) p( \bt^\prime | d_1)p(\bt^\prime | d_2) }{p_{\mathrm{PE}} ^2(\bt^\prime)}  d\bt^\prime }{ \int  \dfrac{p_{\mathrm{obs}}(\bt^\prime) p( \bt^\prime | d_1)}{p_{\mathrm{PE}}(\bt^\prime)} d\bt^\prime \int  \dfrac{p_{\mathrm{obs}}(\bt^\prime) p( \bt^\prime | d_2)}{p_{\mathrm{PE}}(\bt^\prime)} d\bt^\prime},
\end{aligned}
\end{equation}
where $p(\bt^\prime|d_1)$ and $p(\bt^\prime|d_2)$ are the shared posterior parameters (all other detector-frame parameters except effective luminosity distance, arrival times, and phase) of both images, $p_{\mathrm{obs}}$ is the observed population probability and $p_{\mathrm{PE}}$ is the simple prior used in GW parameter estimation.  
As a result, under the lensed hypothesis, all the images come from the same astrophysical source. In the meantime, under the unlensed hypothesis, they come from two different astrophysical sources, resulting in different overall prior volumes (one population term $p_\mathrm{obs}$ in the nominator and two in the denominator). 
Overall the Bayes factor is proportional to the prior volume (inversely proportional to the population prior), meaning that events with similar parameters are more likely to be lensed if the population probability is low.

Interpreting the results of the Bayes factor requires some care, because marginalizing over some of the parameters will reduce the discriminatory power of the Bayes factor; a moderate Bayes factor could become more significant if a full joint analysis was performed. 
This would primarily affect our ability to confidently detect strong lensing.
However, in this study, we will not focus on detectability. 
Instead, we focus on the effect of the population model on the Bayes factor, which depends only on the parameters that enter the BBH population model, masses which we include in the Bayes factor. 
Because of this, the Bayes factors should be interpreted as effectively rank-ordering of the candidates, and not as an absolute indicator of detection (for a detection, computation of the posterior odds would be required). 
To obtain the Bayes factors, we need both the posterior samples from a parameter estimation run and the BBH population probability reconstructed with selection effects (Sec.~\ref{sec:population}).  

\subsubsection{Posterior generation}
As discussed above, the posterior overlap analysis consists of comparing the posterior probability distribution for the parameters of two strong lensing candidates (two different images of the same event). 
To this end, we need to generate posterior distributions for the various events.

The posterior probability density for a particular parameter depends on two quantities: first, the true value of the considered parameter; second, the specific noise realization that comes with the GW itself. 
In particular, in the absence of noise, and assuming a Gaussian probability density function, the posterior probability density function would peak at the true value. 
However, due to noise fluctuations, the peak is typically shifted. 
More specifically, the distribution of the peaks follows a normal distribution centred around the mean value~\citep[e.g.][]{Calskan2022LensingWaves}. 

In this limit, for a generic parameter $\bt^\prime$ and the strain data $d_i=h_i+n_i$ around the $i$th candidate strong GW lensing image, we can write
\begin{align}
    p(\bt^\prime|d_i) = f(\bt^\prime|\hat \bt^\prime, n_i)\,,
\end{align}
where $\hat \bt^\prime$ is the set of true parameters and $f(\bt^\prime|\hat \bt^\prime, n_i)$ is a Gaussian distribution with a mean value drawn from a normal distribution centered around the true value. 
If we consider the case in which two different signals are two images of the same gravitational event and choosing $\bt^\prime$ among the parameters which are not affected by the gravitational lensing (all detector-frame parameters except effective luminosity distances, arrival times, and overall phases), the differences among the two posterior distributions are due to the different noise realizations:
\begin{align}
    p(\bt^\prime|d_1)& = f(\bt^\prime|\hat \bt^\prime, n_1)\\
    p(\bt^\prime|d_2)&= f(\bt^\prime|\hat \bt^\prime, n_2)\,.
\end{align}
On the other hand, if $\bt^\prime$ is affected by the gravitational lensing, the detector will see the two events as if they had two different true values for the same parameter. 
Therefore,
\begin{align}
    p(\bt^\prime|d_1) &= f(\bt^\prime|\hat \bt^\prime_1, n_1)\\
    p(\bt^\prime|d_2) &= f(\bt^\prime|\hat \bt^\prime_2, n_2)\,.
\end{align}
In this case, the two distributions are not expected to overlap.
Since the frequency evolution depends on the detector-frame masses $m_i^z = (1+z)m_i^\mathrm{s}$, where $m_1^\mathrm{s}$ denotes the source-frame masses, and the evolution is not affected by the gravitational lensing, the detector-frame masses are expected to be the same between different GW images and unbiased by strong lensing. 

In this study, we only consider detector-frame masses components for posterior overlap analysis. Instead of generating the $m_1^z, m_2^z$ posteriors, we generate $m_1^z, q$ posteriors for lensed event pairs according to the recipe in Appendix ~\ref{A:mockpost} for more efficient Bayes factor computation. 

\section{Population}\label{sec:population}

We aim to investigate if careless use of a parametric population model or a clearly incorrect model could bias strong lensing detections and if these biases can be mitigated by the use of flexible nonparametric models. 
To perform the study, we will use four population models:
\begin{enumerate}[(i)]
\item a benchmark model that corresponds to the population model used in the mock catalog, \textit{the correct model};
\item a nonparametric model, which is used to reconstruct the underlying model encoded in the available data without assumptions.
\item a physically motivated population model that differs from the benchmark model, the \emph{wrong model};
\item a uniform population model; 
\end{enumerate}
In the following, we will give the details of the parametric models as well as the general idea of the nonparametric model. 
For the scope of this paper, the relevant astrophysical properties of the BH population are the BH source-frame mass (both primary and secondary) and the source redshift.

\subsection{Benchmark model}
In this section, we describe the population used to generate the mock catalog and thus serve as a benchmark model in this work.
For the benchmark model, the primary and secondary masses are taken from the same distribution, which presents an explicit redshift dependence:
\begin{equation}
    p(m^\mathrm{s}_{1,2},z) = p(m^\mathrm{s}_{1,2}|z)p(z)\,.
\end{equation}
The redshift evolution function, $p(z)$, is motivated by analytical fits to the star-formation rate density, similarly to~\citep{Madau_2014},
\begin{align}\label{eq:redshift_dist}
       p(z) =  1 + (1+z_p)^{-\alpha_z-\beta_z} \frac{(1+z)^{\alpha_z}}{1+\qty(\frac{1+z}{1+z_p})^{\alpha_z+\beta_z}}\,,
\end{align}
where $\alpha_{\rm z}=2.7$, $\beta_{\rm z}=2.9$ and $z_p=1.9$.

On the other hand, the source-frame mass distribution $p(m^\mathrm{s}|z)$ is modeled after the PP model: in this model, the power law accounts for the initial mass function (IMF), as in \citep{salpeter1955}, while the Gaussian distribution models a pileup around $45 M_\odot$ that is expected to be due to certain processes---e.g., the pulsational pair-instability supernova (PPISN) process---that some classes of stars (e.g., stars with helium core masses in the range of $\sim 30-50 M_\odot$) may undergo during their life \citep{PPISNe}. The main difference between the mass distribution we adopted in this work and the PP model used in \citep{o3pop} is that we allow for a redshift dependence in the mass function. In particular, $p(m^\mathrm{s}|z)$ reads
\begin{multline}
   P(m_{\rm s}| z)=(1-w(z))\times \mathcal{C} m_{\rm s}^{-2} +\\+ w(z)\mathcal{N}(m_{\rm s}| f_m(z))\,,
\end{multline}
where $\mathcal C$ is the normalization constant  of the mass distribution, $w(z)$ is the redshift-dependent weight,
\begin{align}
    w(z) = w_0\times \qty(1+\frac{z}{z_{\rm max}})\,,
\end{align}
with $w_0\leq 0.15$, and $f_m(z)$ is the standard deviation of the Gaussian distribution
\begin{align}
    f_m(z) = m_0 + m_1\frac{z}{z_{\mathrm{max}}}\,,
\end{align} 
with parameters $m_0 = 55$, $m_1 = 20$ and $z_{\mathrm{max}}=1.3$. 

\subsection{Nonparametric population model}

In this subsection, we will briefly introduce (H)DPGMM, the nonparametric model we use as an agnostic population model. 
We refer the interested reader to \citet{hdpgmm} and references therein, where the model is introduced, and to \citet{figaro}, where \textsc{figaro} \footnote{Publicly available at \url{https://github.com/sterinaldi/} and via \texttt{pip}.}---the implementation we use in this paper to infer the population distribution---is presented for more details.

Nonparametric models are a specific class of models that are able to approximate arbitrary probability densities making use of an infinite number of parameters. In this context, the commonly used name \emph{nonparametric} might be misleading: in fact, these models do have parameters, but they are merely used to describe the shape of the distribution and have no direct connection with the physics governing the processes from which the underlying distribution arises.
(H)DPGMM is a model based on the Dirichlet process Gaussian mixture model (or DPGMM) that is designed to infer probability densities given a catalog of posterior samples for a number of individual GW events. 
The idea behind this model is to approximate an arbitrary probability density with an infinite weighted sum of multivariate Gaussian distributions:
\begin{equation}
    f(m,q,z) \sim \sum^{\infty}_i w_i N(m,q,z|\mu_i,\sigma^2_i)\,.
\end{equation}
The potentially countably infinite parameters\footnote{In any practical implementation of this model, the number of components will be capped at the available number of observations.} of the (H)DPGMM model can be inferred making use of the observations drawn from the underlying distribution $f(m,q,z)$ marginalizing over the individual events masses and redshift---for the specific likelihood, we refer the reader to Sec.~2 and~3 of \citet{hdpgmm}, in particular Eqs.~(39), (41), and~(42).

In this work we use \textsc{figaro}, an implementation of the Gibbs sampling scheme designed to explore the (H)DPGMM model. This algorithm takes as inputs the GW posterior samples catalog and draws multiple vectors of weights, means, and variances, potentially with different lengths as determined by the Dirichlet distribution used as a prior. Each of these vectors correspond to a realization of the (H)DPGMM.

We apply this framework to the problem of inferring the primary mass and the mass ratio of the primary and secondary mass from GW observations. The outcome of this model is a phenomenological distribution that describes the observed events.

\subsection{Wrong model}

The \emph{wrong} population model is a physically inspired population model that does not properly capture the features which are present in the true underlying distribution of the data. 
We included this model to show that biases might arise from wrong population assumptions.

In particular, the detector-frame primary mass distribution and the mass ratio $q = m_2/m_1$ follow power-laws:
\begin{align} \label{eq: pl m1z model}
    p(m_{1,z}) \propto m_{1,z}^{X}\,,
\end{align}
\begin{align}\label{eq: pl q model}
     p(q) \propto q^{Y}\,,
\end{align}
where the power-law indices $X,Y$ can be inferred from the observed catalog.

\subsection{Uniform model}
The uniform model prescripts a uniform distribution on all the involved event parameters. This population model is used in literature---for instance in \citep{Diego2021EvidenceLIGO-Virgo}---to express agnosticism toward the population distribution and to keep it as conservative as possible. We believe, however, that such a choice is incorrect and leads to bias in the Bayes factor computation: in a nutshell, the posterior overlap analysis compares the probability of observing two events that \emph{may look the same} by chance with the probability of having it caused by gravitational lensing. By assuming a flat population, one is neglecting the possibility of having regions in the parameter space that are more likely to be observed than others. Such choice, therefore, artificially enhances the value of the Bayes factor, leading to a significantly higher chance of misclassifying an unlensed event as lensed.
In order to demonstrate the dramatic effects that this incorrect choice can have, we included this model in our analysis.

\subsection{Selection effects} \label{sec:selection fuction}
GWs are detected using the so-called \emph{online search pipelines}, such as \textsc{MBTA} \citep{Aubin:2020goo}, \textsc{pyCBC} \citep{Usman:2015kfa}, \textsc{GstLAL} \citep{Sachdev:2019vvd}, and \textsc{cWB} \citep{Klimenko:2015ypf}: these pipelines release triggers, using the signal-to-noise ratio (SNR) $\rho$ as detection statistics. Whereas the SNR exceeds a threshold value $\rho_\mathrm{th}$, a new event candidate is identified. The SNR is not the sole detection statistics employed to determine whether a stretch of data contains an event, it is accompanied with the false alarm rate (FAR) and the probability of the astrophysical origin of the event $p_\mathrm{astro}$~\citep{poprate_willfarr,poprate_Kapadia_2020}.

In principle, all these quantities play a role in the promotion of an event candidate to confident detection: however, due to the fact that computing the FAR and $p_\mathrm{astro}$ for an event is a computationally expensive task, in what follows we will consider SNR alone.

Each event will produce a different SNR in the detector depending on two factors: its parameters (both intrinsic and extrinsic) and the noise realization surrounding the event. Due to the stochastic nature of the noise, the SNR of a specific event will be a realization of a stochastic process, whose expected value is determined by the event's parameters $\bt$, $\hat\rho(\bt)$.

The detection probability of an event follows the probability of observing $\rho$ greater than $\rho_\mathrm{th}$. According to \citep{allen:2012}, $\rho$ is normally distributed with unitary variance
\begin{equation}
    \rho \sim N(\rho|\hat\rho(\bt),1)\,.
\end{equation}
Therefore, $p_\mathrm{det}$ for a set of parameters $\bt$ is given by
\begin{equation}
    p_\mathrm{det}(\bt) = 1- \int_0^{\rho_\mathrm{th}} N(\rho|\hat\rho(\bt),1) \dd\rho\,,
\end{equation}
where $p_\mathrm{det}$ is the \emph{selection function}.
Concerning $\hat\rho(\bt)$, we relied on the \textsc{python} code \textsc{observation-bias-gw}\footnote{Publicly available at \url{https://github.com/dveske/observation-bias-gw}} \citep{veske:2021}, which provides a script to compute the expected SNR given the primary and secondary mass and the redshift of the source. The angular parameters of the source (sky position, inclination, and polarization) are marginalized out while computing the expected SNR.

\subsection{Mock data catalog} \label{sec:mock data catalog}

We simulated an observed GW catalog composed of 1519 events using the benchmark population model with the selection effect.
Then we sample $m_1^z$ and $q$ posteriors for each event according to the recipe Appendix \ref{A:mockpost}.
The number of events included in the catalog is a middle ground between the number of events that would be needed to have a precise reconstruction with the nonparametric model\footnote{Preliminary studies suggest that we would need tens of thousands of events.
}.
Once the mock posteriors are generated, we perform a hierarchical population inference to get the (H)DPGMM and the power-law model.

In order to compute the Bayes factor for lensed pairs, we need the posteriors for lensed GW pairs.
We repeat the previous process to simulate another astrophysical catalog.
Then for each astrophysical source, we generate two magnification factors using Eq.~[\ref{eq:magnification_marginalised},~\ref{eq:magnification_correlated}] for two images. 
We apply Eq.~[\ref{eq:magnified luminosity distance}] to calculate the effective luminosity distance for each image. 
As a result, the images from the same source have the same $m_1^z$ and $q$ but different luminosity distance.
After that, we calculate the SNR and sample $m_1^z$ and $q$ posteriors for each image separately.

\section{Results}\label{sec:results}

\subsection{Population inference}
\begin{figure*}
    \centering
    \includegraphics[scale = 0.4]{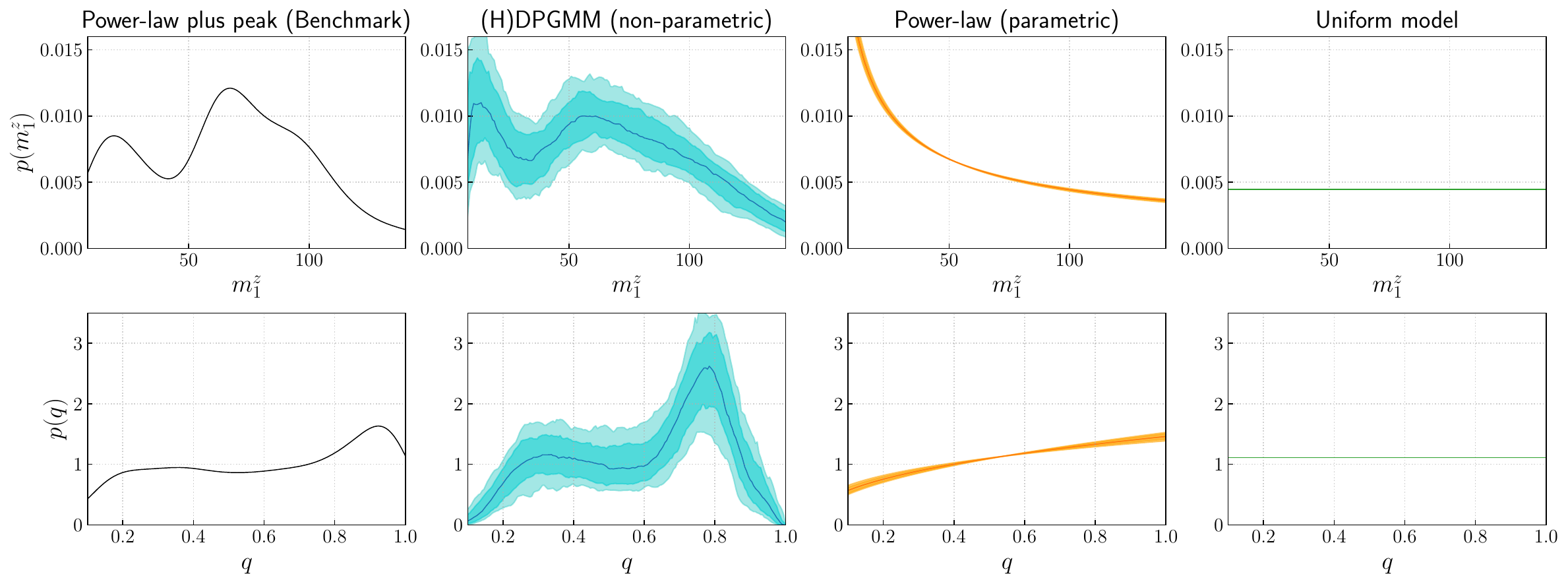}
    \caption{
    Population models for observed distribution. 
    Moving from left to right, the columns correspond to the benchmark model, the inferred (H)DPGMM model, the inferred power-law model, and the uniform model, respectively. 
    The first row represents the marginalized distribution of redshifted primary mass, while the second row represents the marginalized distribution of mass ratio.
    }
    \label{fig:obs_pop_models}
\end{figure*}

We perform the population inference to obtain $p_\mathrm{obs}(m_1^z,q)$ for the (H)DPGMM using the mock observed catalog prepared in Sec.~\ref{sec:mock data catalog}.
For the power-law model, we infer the power-law indices $X, Y$ in Eq.~[\ref{eq: pl m1z model},\ref{eq: pl q model}] using the mock observed catalog to obtain $p_\mathrm{obs}(m_1^z,q)$.
The power-law indices are $X=0.61\pm 0.02, Y=-0.41\pm 0.05$ with one standard deviation.
Figure~\ref{fig:obs_pop_models} shows the marginalized $m_1^z$ and $q$ observed distribution for all population models.

Now, we have 4 population models for the observed distribution:
\begin{enumerate}[(i)
]
\item a benchmark model with the high-dimensional Gaussian peak, which provides the most optimal inference;
\item an inferred (H)DPGMM, which makes fewer model assumptions and could be used in future lensing analyzes; 
\item an inferred power-law model, which assumes the primary mass and mass ratio follow power-law relation when we do not have enough data from observation;
\item a uniform prior, which assumes all the parameters follow a uniform distribution.
\end{enumerate}
We use these four models to evaluate the Bayes factor in favor of strong lensing, with the purpose of investigating the population model's impact on lensed detections. 

\subsection{Population model effects on Bayes factor}
\begin{figure}
    \centering
    \includegraphics[width=\columnwidth]{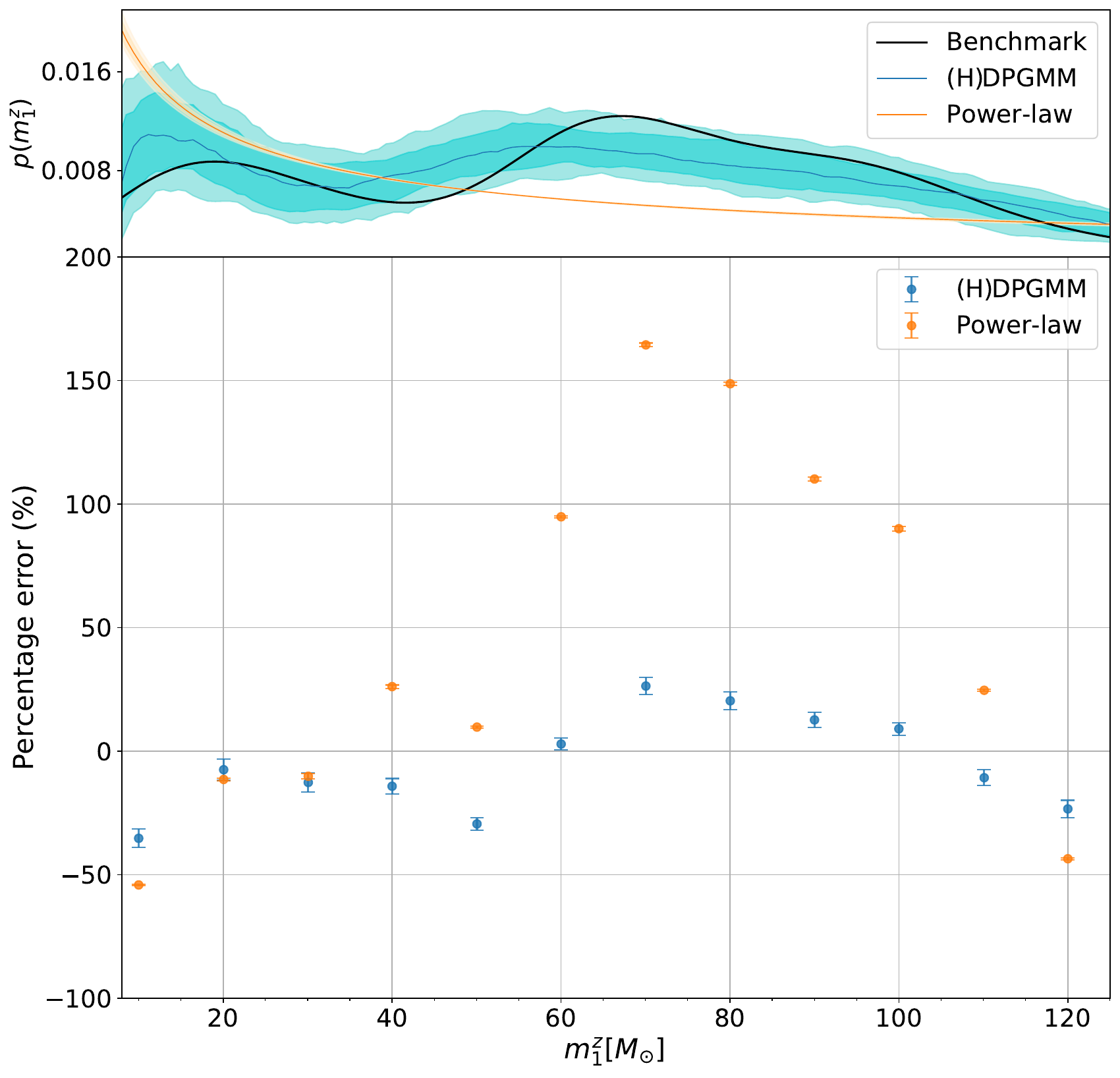}
    \caption{
    The percentage error of lensed pair Bayes factors $\blu$ on $m_1^z$ posterior overlap using different marginalized $m_1^z$ observed population models. 
    The top panel shows the marginalized $m_1^z$ observed distribution from the benchmark model (black), (H)DPGMM (blue) with 68\% and 95\% confidence intervals, and the power-law model (orange). 
    The bottom panel shows the percentage difference of the Bayes factors using (H)DPGMM (blue), power-law model (orange) compare to the benchmark model with 1 standard deviation. 
    The Bayes factors using (H)DPGMM give smaller error than using the power-law model near the Gaussian peak at \(70 \, M_\odot\).}
    \label{fig:strong_lensing_bayes_factors-1D-obs}
\end{figure}
We first evaluate the Bayes factor using only $m_1^z$ posteriors of lensed pairs and marginalized $m_1^z$ observed population model. 

We randomly select 1 event pairs for each $10 M_\odot$ interval from the lensed catalog to perform the Bayes factor calculation for better illustration.
In the upper panel of Fig.~\ref{fig:strong_lensing_bayes_factors-1D-obs}, we show the comparison of benchmark, (H)DPGMM, and power-law models. 
The (H)DPGMM agrees with the benchmark model within the 68\% confidence interval while the power-law model does not. 
As a result, the percentage errors for lensed pair Bayes factors between using (H)DPGMM and benchmark model (blue) are smaller than those between using the power-law model and benchmark model (orange)(Fig.~\ref{fig:strong_lensing_bayes_factors-1D-obs} lower panel). 
The Bayes factors calculated using the (H)DPGMM exhibit an error of approximately \(\pm 30\%\) at 1 standard deviation, whereas those using the power-law model show a larger percentage error near the Gaussian peak at \(70 \, M_\odot\). 
The power-law model exhibits a lower probability compared to the benchmark model and tends to overestimate the Bayes factors, favoring the likelihood of a lensed pair when two observed events have similar event posteriors. 
Note that the Bayes factors computed using the (H)DPGMM have larger error bars compared to those obtained from the power-law model, due to the larger uncertainty in the inferred population model.
This result highlights the impact of the population model on the Bayes factor. 
A parametric model that fails to adequately capture the underlying structure of the population introduces bias into the Bayes factor.
\begin{figure}
    \centering
    \includegraphics[width=\columnwidth]{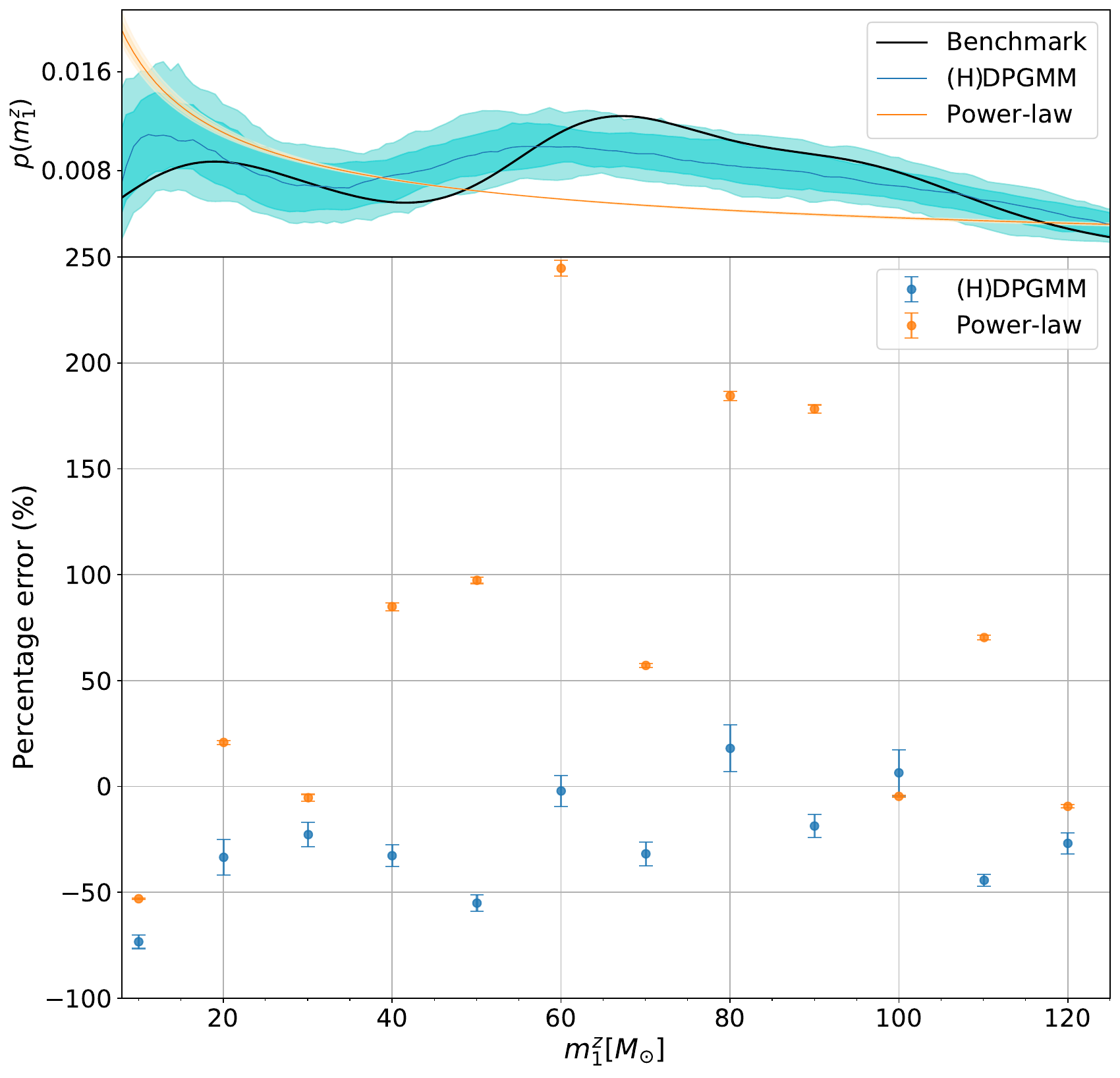}
    \caption{
    The percentage error of lensed pair Bayes factors $\blu$ on $m_1^z,q$ posterior overlap using different observed population models. 
    The top panel shows the marginalized $m_1^z$ observed distribution from the benchmark model (black), (H)DPGMM (blue) with 68\% and 95\% confidence intervals, and the power-law model (orange). 
    The bottom panel shows the percentage difference of the Bayes factors using (H)DPGMM (blue), power-law model (orange) compare to the benchmark model with 1 standard deviation.
    The Bayes factors using (H)DPGMM give $\lesssim 50\%$ error which is smaller than using the power-law model for most of the lensed pairs.}
    \label{fig:strong_lensing_bayes_factors-2D-obs}
\end{figure}

Next, we evaluate the Bayes factor using both $m_1^z$ and $q$ posteriors of lensed pairs to see the error propagation in higher dimensions.
As shown in Fig.~\ref{fig:strong_lensing_bayes_factors-2D-obs}, the percentage error for using (H)DPGMM remains \(\sim \pm 30\%\) while those using power-law model exasperated near the Gaussian peak at \(70 \, M_\odot\).
\subsection{Prior assumptions on Bayes factor}
\begin{figure}
    \centering
    \includegraphics[width=\columnwidth]{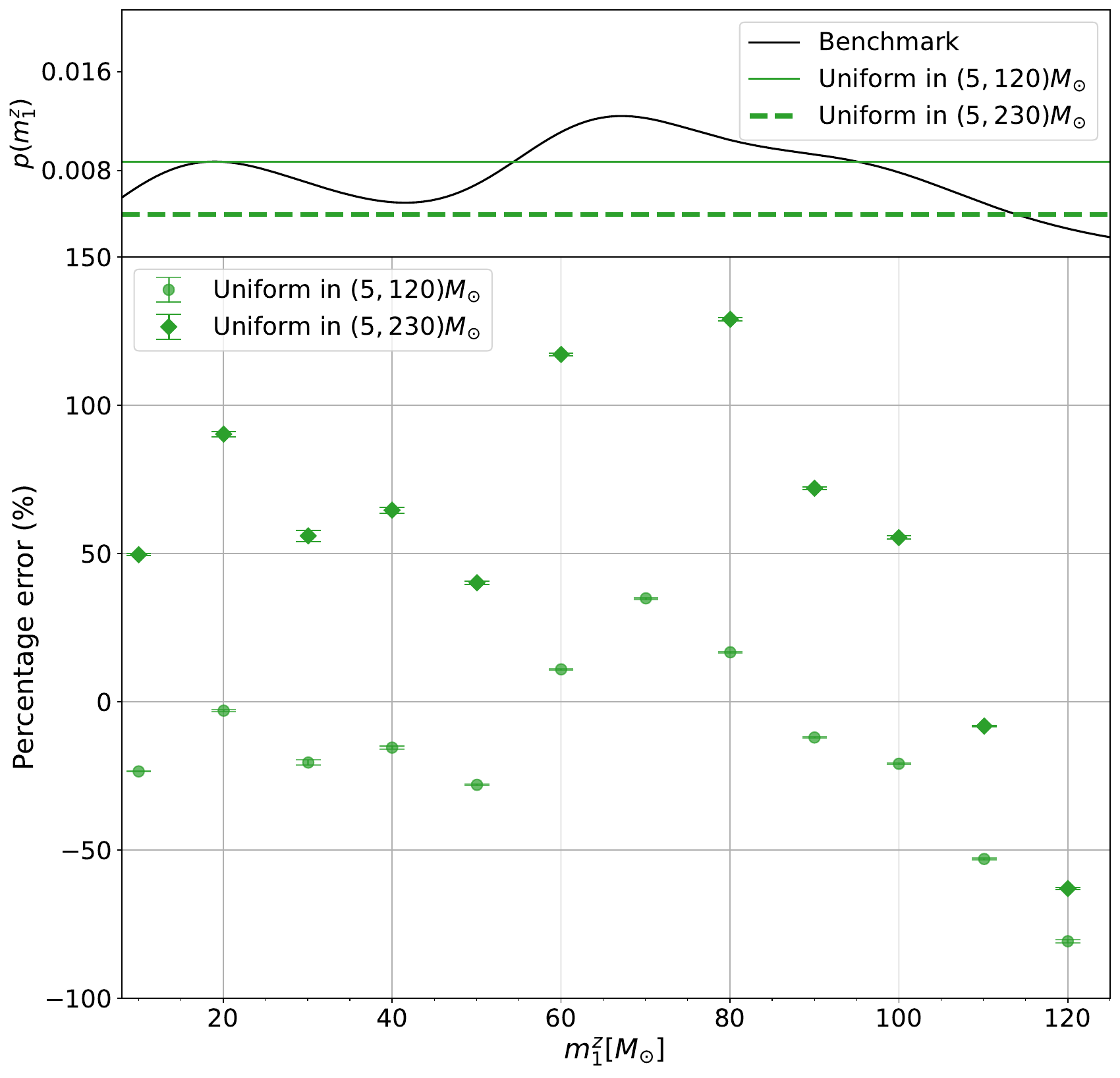}
    \caption{
    The percentage error of lensed pair Bayes factors $\blu$ on $m_1^z$ posterior overlap using the uniform model with different ranges compared to the benchmark observed population model.
    The top panel shows the marginalized $m_1^z$ observed distribution from the benchmark model (black), the uniform model in $(5,120)M_\odot$ (green solid line), and the uniform model in $(5,230)M_\odot$ (green dashed line). 
    The bottom panel shows the percentage difference of the Bayes factors using a uniform model in $(5,120)M_\odot$ (green circle), a uniform model in $(5,230)M_\odot$ (green diamond) compare to using the benchmark model with 1 standard deviation. 
    Using a uniform model bias the Bayes factor.
    }
    \label{fig:strong_lensing_bayes_factors_uniform_prior}
\end{figure}
However, what about the effect of the population model on existing detections? \citep{Diego2021EvidenceLIGO-Virgo} advocated that the Bayes factor evaluated using a uniform prior is less biased than a $\blu$ utilizing a prior informed by population analyzes. 
In Fig.~\ref{fig:strong_lensing_bayes_factors_uniform_prior}, we consider uniform $m_1^z$ prior compared to the benchmark observed model with different ranges.  
The use of a uniform prior leads to overestimation or underestimation of $\blu$ depending on the chosen $m_1^z$ range. Assuming a uniform prior can introduce significant bias because it lacks any features that align with the true population distribution. 
The population probability is heavily influenced by the volume of the parameter space, with larger parameter spaces yielding smaller probabilities under a uniform prior. 
This makes the Bayes factor using a uniform prior arbitrary depending on the range we set and gives inaccurate calculation. 
On the other hand, the result may be counterintuitive because a uniform prior is typically assumed to yield conservative results. 
However, the strong Bayes factor contrasts the probability that two events are strongly lensed against astrophysical coincidence, which depends sensitively on the prior knowledge of the BBH population. 
Indeed, a uniform population model would produce very dissimilar BBHs and therefore highly dissimilar waveforms. 
Thus, under this model, it would be predicted that overlap in two waveforms (and thus the probability of lensing mimickers due to astrophysical coincidence) is unlikely, giving false credence to the strong lensing hypothesis. 
In contrast, a more realistic population model would place greater weight on the astrophysical coincidence. 
Therefore, using a uniform mass prior does not better discriminate strongly lensed events than models that presume more about the population of binaries and lenses. 
Instead, using a uniform mass prior can easily result in a number of false detections of strong lensing~(See Section~\ref{sec:ROC}).

Other than assuming a uniform population prior, if we assume the prior used in parameter estimation to be the same as the observed population (i.e., $p_\mathrm{PE}=p_\mathrm{obs}$), the Bayes factor (Eq.~[\ref{eq:blu formula}]) will reduce to 
\begin{equation} \label{eq:blu with assumption}
\begin{aligned} 
\blus = \int  \frac{p(\bt|d_1)p(\bt|d_2)}{p_\mathrm{obs} (\bt)} d\bt,
\end{aligned}
\end{equation}
where $\blus$ denotes the Bayes factor with such an assumption. 
The Bayes factor (Eq.~[\ref{eq:blu with assumption}]) is often used in various GW lensing literature but it does not correctly account for the population probability. 
We compare the Bayes factor difference using Eq.~[\ref{eq:blu formula}] and Eq.~[\ref{eq:blu with assumption}] with the benchmark population model.
\begin{figure}
    \centering
    \includegraphics[width=\columnwidth]{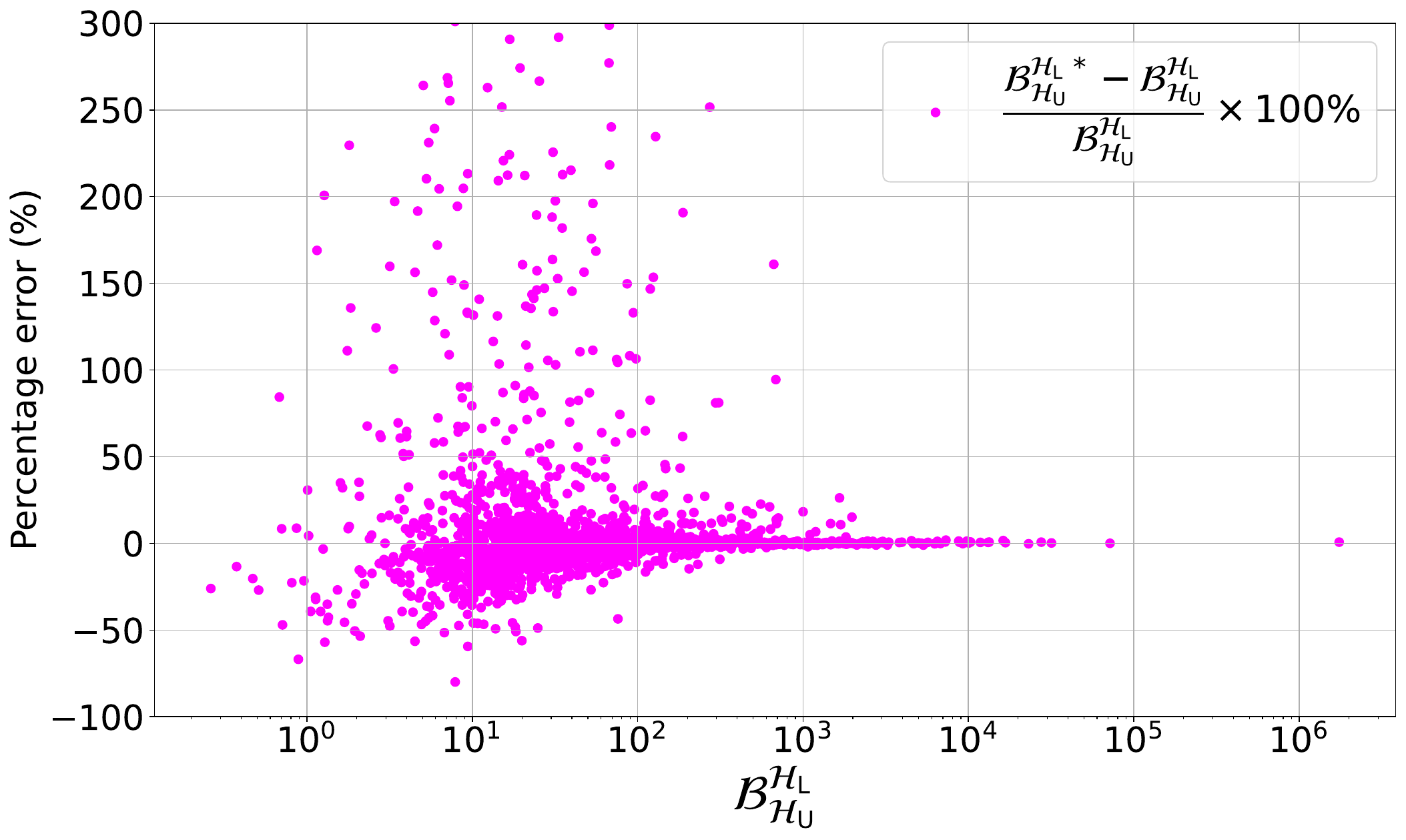}
    \caption{
    The percentage error of lensed pair Bayes factors on $m_1^z, q$ posterior overlap when assuming prior used in parameter estimation to be the same as the observed population ($p_\mathrm{PE}=p_\mathrm{obs}$) versus without such an assumption ($\blus$ versus $\blu$). 
    Bayes factors by assuming $p_\mathrm{PE}=p_\mathrm{obs}$ give stronger bias with smaller $\blu$ magnitude.
    }
\label{fig:strong_lensing_bayes_factors_different_formula}
\end{figure}
Figure~\ref{fig:strong_lensing_bayes_factors_different_formula} shows the percentage difference between the Bayes factors with and without assuming $p_\mathrm{PE}=p_\mathrm{obs}$, which denote by $\blus$ and $\blu$ respectively.
The $\blus$ is biased stronger with a smaller $\blu$.
As the Bayes factor is inversely proportional to $p_\mathrm{obs}$, smaller $p_\mathrm{obs}$ gives a larger Bayes factor and a larger discrepancy between $\blu$ and $\blus$.
It is critical to properly assess the effect of the population model in low-probability regions since two events with similar parameters in such regions are very likely to be lensed pair than astrophysical coincidence. 

\subsection{Detection performance}\label{sec:ROC}

In order to understand how the bias in the Bayes factor affects the strong lensing detection which is essentially a classification problem using the Bayes factor, we plot the receiver operating characteristic (ROC) curves. 
We use 2362 lensed pairs and 2415 unlensed pair to generate the ROC curves by showing the true positive rate (TPR) and false positive rate (FPR), where TPR is the portion of lensed pairs classified correctly as lensed detection and FPR is the portion of unlensed pairs classified wrongly as lensed detection using Bayes factor. 

\begin{figure}
    \centering
    \includegraphics[width=\columnwidth]{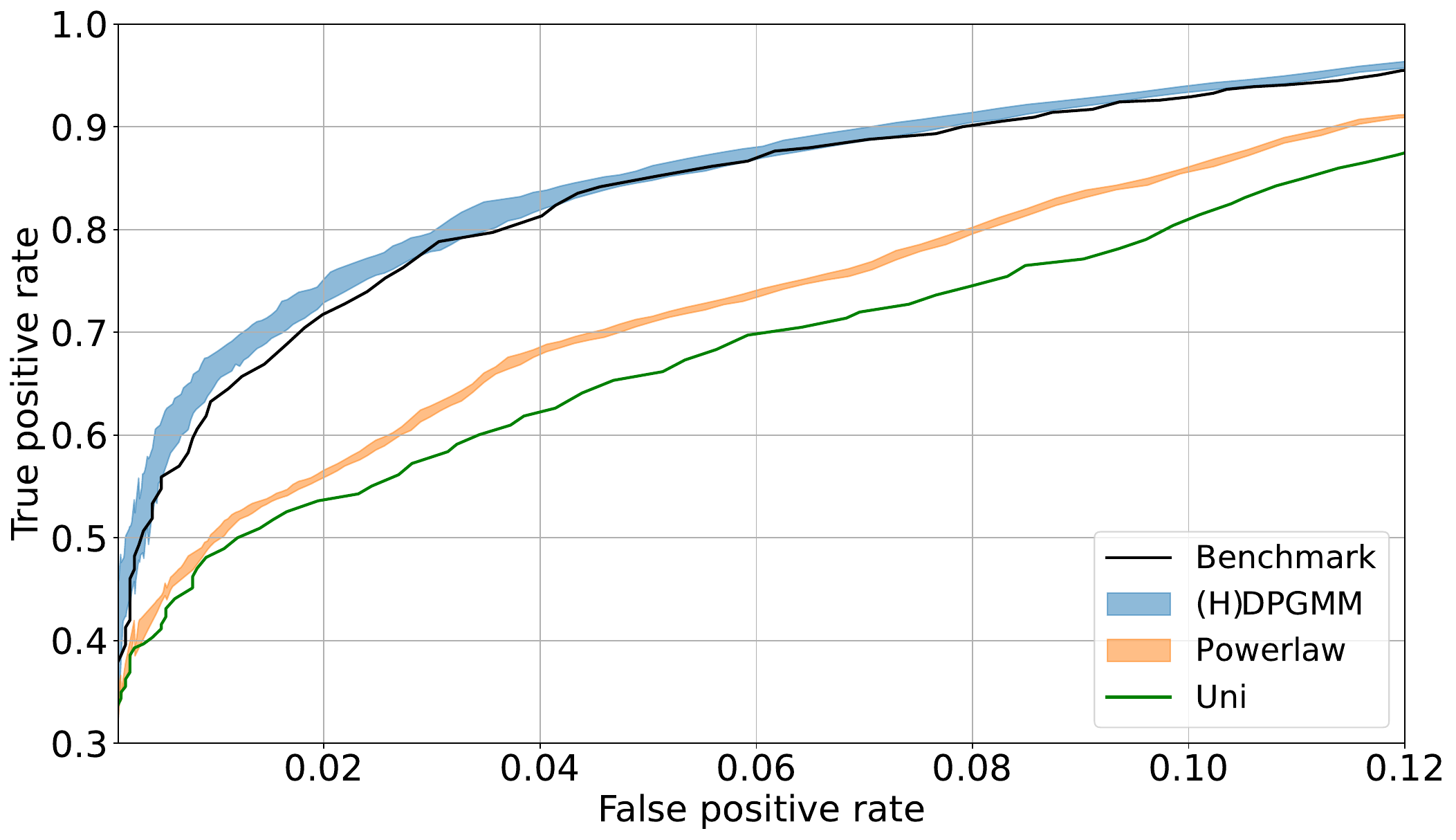}
    \caption{
    Receiver operating characteristic (ROC) curves of strong lensing classification using benchmark population model (black), (H)DPGMM (blue), power-law model (orange), and uniform model (green).
    The ROC curve using (H)DPGMM gives the most comparable performance to the benchmark population model. 
    }
    \label{fig:ROC}
\end{figure}
In Fig.~\ref{fig:ROC}, the ROC curve using the benchmark model (black) shows the best performance on strong lensing detection.
The ROC curve using the (H)DPGMM (blue) gives the most comparable performance to the benchmark ROC curve when compared to the ROC curves using the power-law (orange) and uniform model (green).
This again demonstrates nonparametric population model is more suitable for strong lensing detections.

\begin{figure}
    \centering
    \includegraphics[width=\columnwidth]{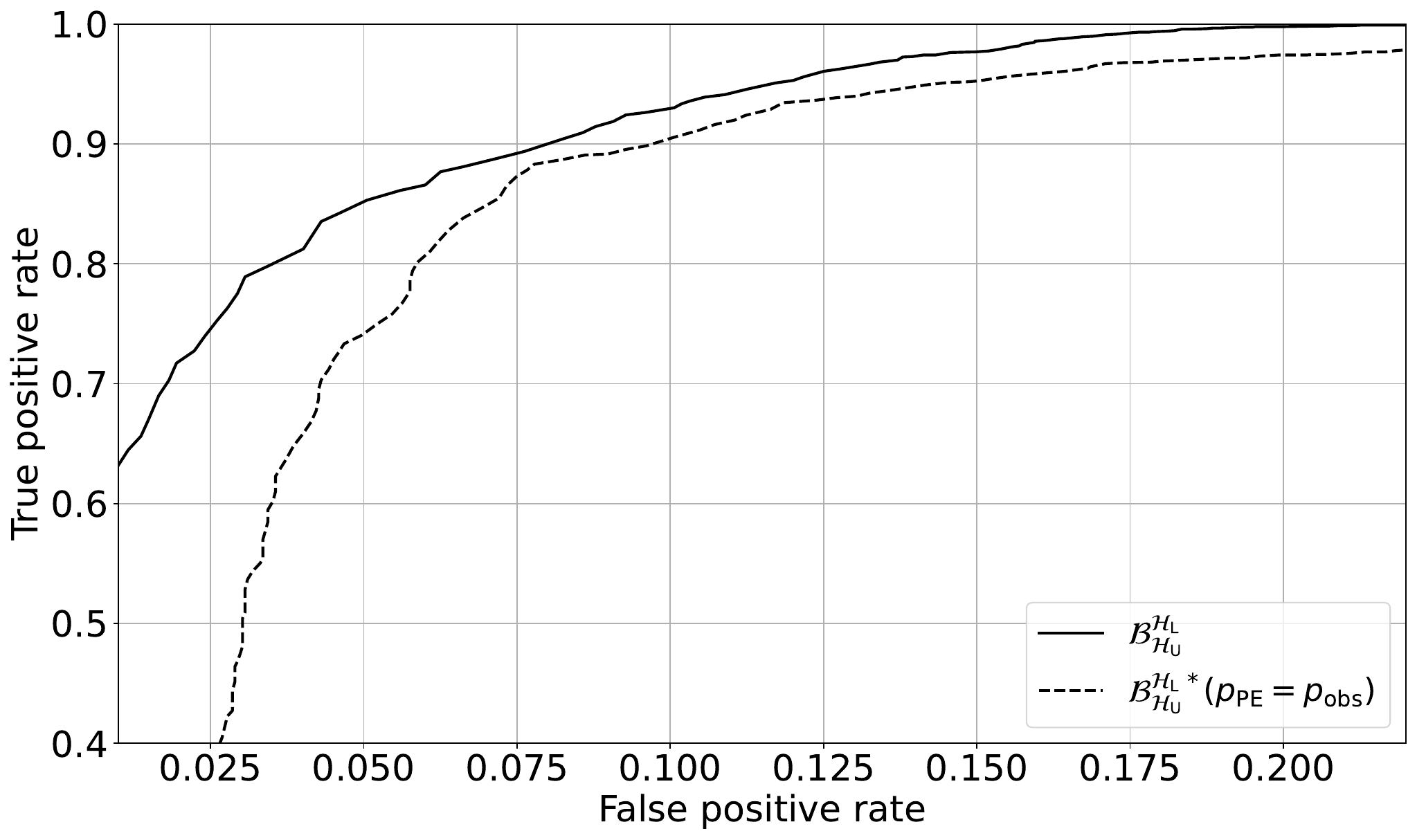}
    \caption{
    Receiver operating characteristic (ROC) curves of strong lensing classification when assuming prior used in parameter estimation to be the same as the population ($p_\mathrm{PE}=p_\mathrm{obs}$) versus without such an assumption ($\blus$ versus $\blu$). 
    ROC curve without such assumption gives better performance for the classification.
    }
\label{fig:ROC_different_formula}
\end{figure}
Figure~\ref{fig:ROC_different_formula} shows the ROC curve comparison on utilizing/not utilizing the assumption $p_\mathrm{PE}=p_\mathrm{obs}$.
The ROC performance is better if such an assumption is not used, implying that Eq.~[\ref{eq:blu formula}] should be used for strong lensing detection. 

Indeed, we advocate for the use of the corrected Bayes factor formula for the posterior overlap analyzes (Eq.~[\ref{eq:blu formula}]). 
We note that the posterior overlap Bayes factor is not used in the joint parameter estimation Bayes factor~\cite{Liu2020IdentifyingVirgo,Lo2021ASignals,Janquart2021AEvents} and so the discussion on the Bayes factor formulation here does not apply to them.

\section{Discussion and Conclusions}\label{sec:conclusions}
The strong lensing analysis relies on accurate modeling of the population of BBHs and strong lenses: 
an incorrect population model could lead to either inappropriate or missed strong lensing detections.

Here we have investigated the use of nonparametric population models to detect strong lensed events.
Unlike parametric modelizations based on phenomenological fits, the nonparametric ones can capture more complicated formation scenarios without assuming a functional form \textit{a priori}.
We find that the use of a nonparametric population model as inferred from mock GW catalog is suitable for strong lensing analyzes: 
while the Bayes factor will end up with broader error bounds than a parametric model that appropriately captures the population, the Bayes factor is less biased by prior assumptions. 

Nevertheless, the convergence of the nonparametric (H)DPGMM observed population model depends on the underlying population. 
Given that the nonparametric model is informed by the data alone and not by the physics behind it, the number of detected events required to properly capture all the features of the underlying distribution depends on the distribution itself---the more complex the distribution, the more events are needed. 
Nonetheless, some preliminary studies that we run on a plausible model based on the 2D $m_1^z-q$ model used in this work show that we need order of thousands of events.

Furthermore,~\citep{Diego2021EvidenceLIGO-Virgo} suggests the analysis with no population prior assumption provides better merger-rate estimation and hence less biased results than the analysis with a PP model in~\citep{TheLIGOScientificCollaboration2021SearchRun}. 
Here, we have shown that a uniform population prior is not an appropriate choice for strong lensing analyzes. 
It strongly biases the result of our analysis on the parameter overlap without considering the merger rate and time-delay distribution.
The uniform prior flattens out the probability and hence gives optimistic estimates of the strong lensing Bayes factor for most of the lensed events, which can lead to more false detections of strong lensing as shown in Fig.~\ref{fig:ROC}.

In addition, the selection effect would not take place in the reconstruction of the population from observed data, which means, we should use the observed population instead of the astrophysical population. 
In Eq.~[\ref{eq:blu formula}], we include the population term accounts for the astrophysical coincidence, however, we cannot see all the astrophysical coincidences because of the selection effect.
Therefore, instead of using an astrophysical population, we should use the observed population when computing the Bayes factor. 
The Bayes factor is used to rank the possibility of lensing among the possible pair seen by our detector instead of all the possible pairs in the Universe. 
Therefore, by using the observed population, the Bayes factor for unlensed pair with similar parameters will be penalized by accounting for the probability of coincidence that is seen by the detector.
This saves lots of effect on reconstructing the astrophysical population from the observed population.

We have employed simplifying assumptions in our modeling of the BBH population, parameter inference, and the strong lensing Bayes factor evaluation. 
Specifically, our parameter inference considers a simplified, parametrized model consistent with current observations, to avoid excessive computational cost. 
Furthermore, our strong lensing Bayes factor is evaluated using the overlap in mass and mass-ratio using the posterior overlap method of~\citealt{Haris2018IdentifyingMergers}, because a full joint parameter estimation evaluating the overlap in the full 15-dimensional BBH parameter space (see~\citep[e.g.][]{Janquart2021AEvents}) could not be used in conjunction with the parametrized model that we use for parameter inference. 
A full analysis employing joint parameter estimation would likely yield higher Bayes factors and would therefore be a better discriminator for whether an event is strongly lensed or not. 
However, with the current methodologies, an extensive joint parameter study is computationally too exhaustive to be applied to all possible pairs of lensed candidates found in a GW catalog. 
Furthermore, the posterior overlap analysis, equipped with the nonparametric population model, can be used to reduce the computational burden of joint PE by prefiltering the pairs that make a good lensed candidate. 
However, we note that including more dimensions in the Bayes factor computation would exasperate the problem: an incorrect population model would lead to even greater discrepancies. 
While these simplified models are adequate for studying the applicability of nonparametric models to strong lensing analyzes, future studies utilizing strong lensing studies, especially on real data should perform joint parameter estimation, which can better quantify if an event is finally strongly lensed or not. 
Indeed, we hope that our work motivates the inclusion of nonparametric models as a core part of strong GW lensing research. 

\section{Acknowledgements}
The authors are grateful to Walter Del Pozzo and Nicola Tamanini for the useful discussions. The authors are thankful to Shasvath J. Kapadia for useful comments on this manuscript. M.T. gratefully acknowledges support from the CNRS/IN2P3 Computing Center (Lyon - France) for providing computing and data-processing resources needed for this work. O.A.H. acknowledges support by grants from the Research Grants Council of Hong Kong (Project No. CUHK 2130822 and 4443086), and the Direct Grant for Research from the Research Committee of The Chinese University of Hong Kong.

\appendix
 \section{Mock posterior}\label{A:mockpost}
 Here we present a simple mock posterior generation for GW events in our mock catalog. 
For each GW event with true event parameter $(m_1^z,q)$, we want to generate $(m_1^{z,\mathrm{obs}},q^{\mathrm{obs}})$ samples that mimic the inference result from parameter estimation using the strain data.

First, due to the fact that noise is included in the real strain data, the maximum likelihood from parameter estimation is not referring to the true event parameter. 
In addition, the noise introduced measurement uncertainties in the parameter estimation.
To include these effects, we first draw a sample $\hat x $ from the true value and then sample the posterior from a distribution characterized by $(\hat x, \sigma_x)$, where $\sigma_x$ is the parameter that takes the measurement uncertainty into account. 

We follow the approach outlined in \cite{mock_posterior}, with some modifications to the mass ratio generation process. 
First, we draw the observed SNR from a Gaussian distribution  
\begin{equation} \label{eq:rho_obs} 
\rho_\mathrm{obs} = \rho + \mathcal{N}(0, 1),
\end{equation}  
where \(\mathcal{N}(\mu, \sigma)\) denotes a normal distribution with mean \(\mu\) and standard deviation \(\sigma\).  

For the detector-frame chirp mass \(\mathcal{M}_z\), we first draw a reference sample $\hat{\mathcal{M}_z}$ by
\begin{equation} \label{eq:mc_ref_generation}
\log \hat{\mathcal{M}_z} = \log \mathcal{M}_z + \mathcal{N}(0, \sigma_m / \rho_\mathrm{obs}),
\end{equation}  
and the detector-frame chirp mass posteriors then are drawn from
\begin{equation} \label{eq:mc_obs_generation}
\log \mathcal{M}_{z, \mathrm{obs}} = \log \hat{\mathcal{M}_z} + \mathcal{N}(0, \sigma_m / \rho_\mathrm{obs}).
\end{equation}  

We generate the mass ratio posterior using a truncated skew-normal distribution. 
The location parameter \(\xi\) is derived from the true mass ratio \(q\)
\begin{equation} \label{eq:skew_normal_loc}
\xi = q - \frac{\sigma_q}{\rho_\mathrm{obs}} \cdot \frac{\alpha}{\sqrt{1 + \alpha^2}},
\end{equation}  
where \(\alpha\) is the skewness parameter. The truncated skew-normal distribution \(\mathcal{SN}^{[0,1]}\) ensures that samples are within \([0, 1]\).  
First, we generate a reference sample
\begin{equation} \label{eq:q_ref_generation}
\hat{q} = \mathcal{SN}^{[0,1]}(q, \sigma_q / \rho_\mathrm{obs}, \alpha),
\end{equation}  
and then draw posterior samples for \(q^{\mathrm{obs}}\) as  
\begin{equation} \label{eq:q_obs_generation}
q_{\mathrm{obs}} = \mathcal{SN}^{[0,1]}(\hat{q}, \sigma_q / \rho_\mathrm{obs}, \alpha).
\end{equation}  
 
The detector-frame primary mass is calculated using 
\begin{equation} \label{eq:m1z_generation}
m_{1,\mathrm{obs}}^z = \frac{M + \sqrt{M^2 - 4 \eta M^2}}{2},
\end{equation}  
where \(M = \mathcal{M}_{z,\mathrm{obs}} / \eta^{3/5}\) and \(\eta = q_{\mathrm{obs}} / (1 + q_{\mathrm{obs}})^2\) is the symmetric mass ratio.  

The measurement uncertainty parameters are fitted using 70 events from GWTC-3  
\[
\sigma_m = 0.11 \cdot \rho_\mathrm{thresh}, \quad \sigma_q = 0.22 \cdot \rho_\mathrm{thresh}, \quad \alpha = -0.0066,
\]  
where the SNR threshold is set to \(\rho_\mathrm{thresh} = 8\). 

By following the steps outlined in Eqs. (\ref{eq:rho_obs})–(\ref{eq:m1z_generation}), we generate posterior samples for each GW event in the mock catalog. These samples represent the likelihood distribution of the observed GW event parameters, effectively mimicking the results of parameter estimation.

\onecolumngrid
\section{Odds Ratio derivation}\label{A:blu_derivation}
Here, we present the derivation of $\blu$ in Eq.~[\ref{eq:blu formula}] following the formalism in \citet{Lo2021ASignals}.
To determine whether a set of gravitational-wave events are lensed or not, we make use of the quantity called the odds ratio
\begin{equation}
\olu = \frac{p(\hl|\bd)}{p(\hu | \bd)} = 
\frac{p(\hl)}{p(\hu)} \frac{p(\bd|\hl)}{p(\bd|\hu)} =
\plu \blu,
\end{equation}
where $\bm{d}=\{d_1,\ldots d_N\}$ is the dataset contains $N$ signals, $\plu$ is the prior odds and $\blu$ is the Bayes factor.
And lensed and unlensed hypotheses are denoted as
\begin{itemize}
    \item $\hl$: The dataset {$\bm{d}$} contains lensed signals from a
    single binary black hole merger event with parameters $\bt^{(1)}=...=\bt^{(N)}$.
    \item $\hu$: The dataset {$\bm{d}$} contains signals from independent binary black hole merger events with parameters $\bt^{(i)}$.
\end{itemize}
\vspace{0.5cm}
Here, we focus on the derivation of the Bayes factor. First, we focus on $p(\bd|\hl)$ in $\blu$. 
We can expand the probability in event parameter $\bt$ space. In addition, we can separate $\bt$ into two independent subsets: $\btu$ and $\bteff$, where
\begin{itemize}
    \item $\btu$: parameters that do not affect by the lensing mechanism, such as masses, and spins.
    \item $\bteff$: parameters that will be affected by the lensing mechanism such as luminosity distance, arrival time, and phase difference.
\end{itemize}
Then we have
\begin{equation}
p(\bd|\hl) = \int p(\bd|\btu,\bteff)p(\btu|\hl)p(\bteff|\hl) d\btu d\bteff,
\end{equation}
where $p(\btu|\hl)$and $p(\bteff|\hl) $ are the priors used for $\btu$and $\bteff$ under lensed hypothesis.
Then, we can rewrite it with the following trick:
\begin{equation} \label{eq:hl evidence}
\begin{aligned}
p(\bd|\hl)  &= \int p(\bd|\btu,\bteff) p(\btu|\hl)p(\bteff|\hl) d\btu d\bteff \\
            &= \int p(\bd|\btu,\bteff) p(\btu|\hl)p(\bteff|\hl) \dfrac{p(\btu|\ho)p(\bteff|\ho)  }{p(\btu|\ho)p(\bteff|\ho)} d\btu d\bteff \\
            &= \int p(\bd|\btu,\bteff) p(\btu|\ho)p(\bteff|\ho) \dfrac{p(\btu|\hl)p(\bteff|\hl) }{p(\btu|\ho)p(\bteff|\ho) } d\btu d\bteff,
\end{aligned}
\end{equation}
where $p(\btu|\ho)$and $p(\bteff|\ho) $ are the priors used for $\btu$and $\bteff$ in parameter estimation.
Here, we can use Bayes' theorem to write the product of the posterior and the evidence as 
\begin{equation} \label{eq:Bayes theorem}
        p( \btu,\bteff | \bd )   p(\bd | \ho)  =p(\bd | \btu, \bteff)  p(\btu|\ho)p(\bteff | \ho).
\end{equation}
We apply Eq.[\ref{eq:Bayes theorem}] on Eq.[\ref{eq:hl evidence}] to replace the first two terms in the integral and it gives:
\begin{equation}
p(\bd|\hl)   = p(\bd | \ho) \int  p( \btu,\bteff | \bd )  \dfrac{p(\btu|\hl)p(\bteff|\hl) }{p(\btu|\ho)p(\bteff|\ho) } d\btu d\bteff.
\end{equation}
Notice that we assumed the signals are not overlapping and the parameter estimations are not correlated for different
events. Therefore, we can write it as
\begin{equation}
\begin{aligned}
p(\bd|\hl)  &= p(\bd|\ho) \int  \dfrac{p(\bteff|\hl) \prod^{N}_i p(\btu^{(i)}|\hl) p( \btu^{(i)},\bteff^{(i)}|d_i )}{\prod_i p(\btu^{(i)}|\ho)p(\bteff^{(i)}|\ho)} d\btu^{(1)}\ldots d\btu^{(N)} d\bteff ,
\end{aligned}
\end{equation}
And we expect $\btu^{(i)}$ to be the same across $N$ events under the lensing hypothesis. Therefore, we have $(N-1)$ Dirac-delta distribution to reduce the dimension of the integral
\begin{equation}
\begin{aligned}
p(\bd|\hl)  &= p(\bd|\ho) \int  \dfrac{p(\btu^{(N)}|\hl)p(\bteff|\hl) \prod^{N}_i p( \btu^{(i)},\bteff^{(i)}|d_i )}{\prod_i p(\btu^{(i)}|\ho)p(\bteff^{(i)}|\ho)} d\btu^{(N)} d\bteff,\,
\bt^{(1)}=...=\bt^{(N)}.
\end{aligned}
\end{equation}

where $p(\bd | \ho)$ is the evidence, the first two terms of the nominator are the population prior, followed by the event posteriors. The denominator is the prior used in the parameter estimation.

We can do the same on the unlensed hypothesis,
\begin{equation}\label{eq:hu evidence}
\begin{aligned} 
p(\bd|\hu)  &= \int p(\bd|\bt) p(\bt|\hu) d\bt
    = \int p(\bd|\bt) p(\bt|\hu) \dfrac{p(\bt|\ho)  }{p(\bt|\ho)} d\bt  
    = p(\bd | \ho) \int p(\bt|\bd)  \dfrac{p(\bt|\hu) }{p(\bt|\ho) } d\bt\\
    &= p(\bd | \ho) \prod_i \int p(\bt^{(i)}|d_i)  \dfrac{p(\bt^{(i)}|\hu) }{p(\bt^{(i)}|\ho) } d\bt^{(i)}.
\end{aligned}
\end{equation}
In conclusion
\begin{equation} \label{eq:blu all parameters}
\begin{aligned} 
\blu    &= \frac{p(\bd|\hl)}{p(\bd|\hu)}
        =\dfrac{\int  \dfrac{p(\btu^{(N)}|\hl)p(\bteff|\hl) \prod^{N}_i p( \btu^{(i)},\bteff^{(i)}|d_i )}{\prod_i p(\btu^{(i)}|\ho)p(\bteff^{(i)}|\ho)} d\btu^{(N)} d\bteff}{\prod_i \int p(\bt^{(i)}|d_i)  \dfrac{p(\bt^{(i)}|\hu) }{p(\bt^{(i)}|\ho) } d\bt^{(i)}}.
\end{aligned}
\end{equation}
For our case, we assume only two images are produced. 
In addition, only masses components are considered (no $\bteff$). 
\begin{equation} \label{eq:blu 2img}
\begin{aligned} 
\blu &=
    \dfrac{\int  \dfrac{p(\bt|\hl)  p( \bt|d_1) p( \bt|d_2) }{p(\bt|\ho)p(\bt|\ho)} d\bt}{ \int p(\bt^{(1)}|d_1)  \dfrac{p(\bt^{(1)}|\hu) }{p(\bt^{(1)}|\ho) } d\bt^{(1)}\int p(\bt^{(2)}|d_2)  \dfrac{p(\bt^{(2)}|\hu) }{p(\bt^{(2)}|\ho) } d\bt^{(2)}}
\end{aligned}
\end{equation}
Also, we assume the population prior under $\hl$ and $\hu$ to be the same since the probability of strong lensing is very low. We denote the astrophysical population prior to be $p_{\mathrm{A}}$ and the simple prior used in \textbf{ } to be $p_{\mathrm{PE}}$. Then the Bayes factor can be expressed as
\begin{equation} 
\begin{aligned} 
\blu    &=\dfrac{\int  \dfrac{p_{\mathrm{A}}(\bt^\prime) p( \bt^\prime | d_1)p(\bt^\prime | d_2) }{p_{\mathrm{PE}} ^2(\bt^\prime)}  d\bt^\prime }{ \int  \dfrac{p_{\mathrm{A}}(\bt^\prime) p( \bt^\prime | d_1)}{p_{\mathrm{PE}}(\bt^\prime)} d\bt^\prime \int  \dfrac{p_{\mathrm{A}}(\bt^\prime) p( \bt^\prime | d_2)}{p_{\mathrm{PE}}(\bt^\prime)} d\bt^\prime},
\end{aligned}
\end{equation}
where $p(\bt^\prime|d_1)$ and $p(\bt^\prime|d_2)$ are the shared posterior parameters of images 1 and 2 respectively. 
If we use a simple uniform prior in the parameter estimation, the formula is simplified to 
\begin{equation} 
\begin{aligned} 
\blu    &=\dfrac{\int  p_{\mathrm{A}}(\bt^\prime) p( \bt^\prime | d_1)p(\bt^\prime | d_2)(\bt^\prime) d\bt^\prime }{ \int  p_{\mathrm{A}}(\bt^\prime) p( \bt^\prime | d_1) d\bt^\prime \int  p_{\mathrm{A}}(\bt^\prime) p( \bt^\prime | d_2) d\bt^\prime}.
\end{aligned}
\end{equation}

\twocolumngrid
\bibliography{bibliography}

\label{lastpage}
\end{document}